\documentclass[12pt,preprint]{aastex}

\begin{document}
\bibliographystyle{unsrt}
%\draft
\title{Spin-down Rate of Pinned Superfluid}
\author{M. Jahan-Miri}
\affil{Department of Physics, Shiraz University, Shiraz 71454,
Iran}
\authoremail{jahan@physics.susc.ac.ir}

\begin{abstract}
The spinning down (up) of a superfluid is associated with a radial
motion of its quantized vortices. In the presence of pinning
barriers against the motion of the vortices, a spin-down may be
still realized through ``random unpinning'' and ``vortex motion,''
as two physically separate processes, as suggested recently. The
spin-down rate of a pinned superfluid is calculated, in this
framework, by directly solving the equation of motion applicable
to only the unpinned moving vortices, at any given time. The
results indicate that the pinned superfluid in the crust of a
neutron star may as well spin down at the same steady-state rate
as the rest of the star, through random unpinning events, while
pinning conditions prevail and the superfluid rotational lag is
smaller than the critical lag value.

\end{abstract}

\section{Introduction}
Spinning down (up) of a superfluid at a given rate is associated
with a corresponding rate of outward (inward) radial motion of its
quantized vortices. If the vortices are subject to pinning, as is
observed in the experiments on superfluid Helium
(\markcite{hed80}Hedge \& Glaberson 1980; \markcite{sch81}Schwarz
1981; \markcite{adam85} Adams, Cieplak \& Glaberson 1985;
\markcite{Ziv02}Zieve \& Donev 2000) and also assumed for the
superfluid in the crust of a neutron star (pinned to the lattice
nuclei) (\markcite{T75}Tsakadze \& Tsakadze 1975;
\markcite{TT}Tsakadze \& Tsakadze 1980; \markcite{A87} Alpar 1987;
\markcite{titi90}Tilley \& Tilley 1990), a spin-down would require
also unpinning of the vortices, in order to become moveable.
Unpinning may be realized by the combined effects due to the
Magnus effect, quantum tunnelling and/or thermal activation.
However, the subsequent {\em radial} motion of the unpinned
vortices (before repinning) is a separate {\em dynamical} process,
subject to their equation of motion, apart from the unpinning
process. This is a view different than that adopted in the model
of ``vortex creep" (\markcite{alet84} Alpar et.~al. 1984;
\markcite{jonAp91}Jones 1991b; \markcite{elb92}Epstein, Link \&
Baym 1992), which envisages the spin-down to occur through quantum
tunnelling {\it between} adjacent pinning sites, at different {\em
radial} distances. A critical discussion of the model of vortex
creep, as well as further justification of the presently adopted
viewpoint, may be found elsewhere (\markcite{MJ05}Jahan-Miri
2005a;\markcite{MJ05b}Jahan-Miri 2005b). The derivation of the
spin-down rate of a superfluid, in presence of random unpinning,
as discussed here, aims to pay due attention to the dynamical role
of the vortex {\em radial} motion. That is, vortex radial motion
accompanies a transfer of the spin-down (-up) torque between the
``container'' and the bulk superfluid, which has to be necessarily
meditated by the {\em moving} (not the stationary {\em pinned})
vortices, as in the absence of any pinning (\markcite{son87}Sonin
1987; \markcite{titi90}Tilley \& Tilley 1990). Nevertheless, there
exist uncertainties in the (micro)physics of vortex motion, as
opposed to the structure of a vortex lattice, as well as in the
theoretical understanding of the pinning/unpinning mechanisms.
Such issues are beyond the scope of the present discussion, and
are dealt with by making justified assumptions. The predicted
general relation here reduces to that reported previously
(\markcite{MJ05a}Jahan-Miri 2005a), as an approximate limiting
case. Moreover, the present calculation is based on a direct
solution of the equation of motion for the (temporarily unpinned
movable) vortices, in contrast to the heuristic arguments used in
\markcite{MJ05a}Jahan-Miri 2005a.

\section{The Spin-down Rate}
Different aspects of the derivation will be first discussed
separately, which will be then put together to infer the spin-down
rate of a superfluid, in the presence of random unpinning of the
vortices.

\subsection{Vortex Dynamics}
The total number density $n_{\rm v}$, per unit area, of the
vortices (pinned and unpinned, as a whole) in a superfluid
rotating at a rate $\Omega_{\rm s}$ is
\begin{eqnarray}
\kappa n_{\rm v} = 2 \Omega_{\rm s},
\end{eqnarray}
where $\vec{\kappa}$ is the vorticity vector of a vortex line
directed along its rotation axis. A given rate $\dot \Omega_{\rm
s}$ of change of the rotation frequency $\Omega_{\rm s}$ of the
superfluid is associated with a (averaged) radial velocity $v_r$
of the vortices:
\begin{eqnarray}
\dot \Omega_{\rm s} &=& - 2 \ { \Omega_{\rm s} \over r} \ {v_r} ,
\end{eqnarray}
where $r$ is the distance from the rotation axis, and $v_r>0$ is
in the outward direction. The vortices move with the local
superfluid velocity except when there is an external force acting
on them. The vortex equation of motion is given as (\markcite{son87}Sonin 1987):
\begin{eqnarray}
        \vec{F}_{\rm ext} + \vec{F}_{\rm M}=0
\end{eqnarray}
where $\vec{F}_{\rm ext}$ is the external force on a vortex, per
unit length, exerted by the environment/container of the
superfluid. The kinematic side of the equation is represented by
the Magnus term $\vec{F}_{\rm M}$, arising from the gradient of
the superfluid kinetic energy, (loosely, referred to as a "force''
exerted by the superfluid on the vortices) and is given, per unit
length of a vortex, as (\markcite{son87}Sonin 1987)
\begin{eqnarray}
\vec{F}_{\rm M} & = &  \rho_{\rm s} \vec{\kappa} \times
                        (\vec{v}_{\rm L} - \vec{v}_{\rm s}),
\end{eqnarray}
where $\rho_{\rm s}$ is the superfluid density, $\vec{v_{\rm s}}$
is the local velocity of the superfluid, and $\vec{v_{\rm L}}$ is
the velocity of the vortex-line. Hence, a {\em radial} motion of a
vortex, associated with a spin-down, has to be accompanied and
indeed driven by a corresponding {\em azimuthal external} force
$F_{\rm ext}$ acting on the {\em moving} vortex, instantaneously;
a preliminary fact, however.

\subsection{Unpinning}
In the presence of vortex pinning, the required unpinning of the
vortices may be normally achieved (in the laboratory cases, and
also in neutron stars) under the influence of the so-called
(radial) Magnus ``force'' (\markcite{adam85} Adams, Cieplak \&
Glaberson 1985; \markcite{alet84} Alpar et.~al. 1984). Given a
rotational lag \( \omega \equiv \Omega_{\rm s} - \Omega_{\rm L}\)
between the rotation frequencies of the superfluid and the
vortices (where $\Omega_{\rm L} = \Omega_{\rm c}$, if vortices are
further assumed to be pinned and co-rotating with the
container/crust), radially directed pinning forces would be
effective, and are balanced out by the corresponding component of
the Magnus term \((F_{\rm M})_r= \rho_{\rm s} \kappa r \omega \),
where \(\omega
>0 \) corresponds to an outward directed $(F_{\rm M})_r$ (Eq.~4).
A critical lag $\omega_{\rm crit}$ is thus defined as the maximum
value of the lag that the available pinning forces can sustain.
For larger assumed values of the lag, $\omega \geq \omega_{\rm
crit}$, however stationary pinning conditions may not be realized,
and the spinning down of the superfluid occurs as in the absence
of any pinning, while {\em all} of the vortices move and are
influenced by the existing external forces, instantaneously.

In contrast, when $\omega< \omega_{\rm crit}$, which is the case
of interest here, vortices might be still released from their
pinning sites, though partially and temporarily, due to other
unpinning mechanisms, say, random unpinning through quantum
tunnelling \&/or thermal activation (\markcite{alet84} Alpar
et.~al. 1984). Vortices unpin randomly, move to new radial
positions under the influence of the external forces, and pin
again; the superfluid spins down accordingly. However, the crucial
distinction, with the above case of complete unpinning due to the
Magnus effect, is that at any given instant, only a {\em fraction}
of the total number of the vortices (ie, the movable unpinned
ones) take part in the transmission of an spin-down torque to the
superfluid. The number density $n_{\rm m}$ of the instantaneously
moving vortices, that should be considered in a calculation of the
superfluid spin-down rate, may be written as
\begin{eqnarray}
n_{\rm m} = \xi \ n_{\rm v},
\end{eqnarray}
where $\xi$ is the fraction of the statistical population of
unpinned vortices, at any given time. The superfluid spin-down
rate, under the assumed condition of $|\omega|< \omega_{\rm crit}$
and random unpinning of its vortices, would be therefore regulated
by the unpinning probability $\xi$, being the weight function for
the instantaneous number of (unpinned) moving vortices. This is in
spite of the fact that the spin frequency of the superfluid would
be still determined by the total number density $n_{\rm v}$ of the
vortices (pinned and unpinned), as in Eq.~1. A determination of
the unpinning probability is subject to theoretical uncertainties,
as discussed by various authors (\markcite{ank64}Anderson \& Kim
1964; \markcite{alet84} Alpar et.~al. 1984;
\markcite{jonAp91}Jones 1991b; \markcite{elb92}Epstein, Link \&
Baym 1992). For definiteness, and comparison, the same
prescription as given in the vortex creep model may be used. An
energy barrier $\Delta E= E_{\rm p}(1- \omega/\omega_{\rm cr})$ is
associated with the pinning potential, per unit length, and $\xi$
is given as (\markcite{alet84} Alpar et.~al. 1984)
\begin{eqnarray}
\xi = \exp{(-\Delta E/ kT)}= \exp{[- \frac{E_{\rm p}}{{\rm k}T}
\frac{\omega_{\rm crit}-\omega}{\omega_{\rm crit}}]},
\end{eqnarray}
where $E_{\rm p}$ is the pinning energy, k is the Boltzmann
constant, and $T$ is the temperature.

\subsection{``External'' Forces}
A spin-down of the superfluid would require, in addition to the
freedom of the vortices to move in the interstitial space, also
the presence of {\em azimuthal external} forces acting on the {\em
unpinned moving} vortices, instantaneously (\S~2.1). This is a
fundamental requirement, irrespective of the nature of the
unpinning mechanism, and also of the presence or absence of any
rotational lag between the superfluid and the vortices. It may be
noted that a treatment of the possible ``pinning'' of the vortices
to the local minima of energy in the interstitial medium is not
addressed here, and only pinning to the localized sites (array of
the nuclei in the crust of a neutron star) is considered, with a
single $\omega_{\rm crit}$ associated to each superfluid layer, as
customary. The external forces on vortices could, in general, be
of a viscous drag or a ``static'' frictional nature
(\markcite{adam85}Adams, Cieplak \& Glaberson 1985;
\markcite{jon91}Jones 1991a). The latter type, associated with the
``pinning'' forces should not be however confused with the role of
pinning forces on the ``stationary'' pinned vortices co-rotating
with the pinning centers. In order for the pinning forces to act
as frictional forces and impart a net torque on the superfluid the
vortices should remain unpinned due to the effect of the Magnus
effect (\markcite{adam85} Adams, Cieplak \& Glaberson 1985). This
requires $|\omega|>\omega_{\rm crit}$ (usually assumed to hold in
any given shell of the star, being a stronger condition than the
actual requirement for each vortex line to unpin with a minimum
relative velocity with respect to the local superfluid), which
means there should be no stationary pinning, hence no random
unpinning, to start with. Therefore, the static frictional forces
are not relevant to the case of interest here, where
$|\omega|<\omega_{\rm crit}$ is assumed. On the other hand, the
viscous drag force depends on the relative azimuthal velocity
$v_{\rm rel}$ between the ``container'' and the {\it unpinned}
vortices, and also on the associated microscopic
velocity-relaxation timescale $\tau_v$ of the vortices. The drag
force $F_{\rm d}$, per unit length, is given, for the case of free
vortices in the absence of pinning, as (\markcite{as88}Alpar \&
Sauls 1988)
\begin{eqnarray}
n_{\rm v} F_{\rm d} & = &  \rho_{\rm c} {v_{\rm rel} \over \tau_v
},
\end{eqnarray}
where $\rho_{\rm c}$ is the effective density of the
``container''. For the superfluid in the crust of a neutron star,
the permeating electron (and phonon) gas co-rotating with the
solid crust exert the drag forces on the vortex cores. The
``container'' in this case would be the ``crust'' which includes
all the other components of the star, apart from the superfluid in
the crust, and consists of the solid lattice, phonons, and the
permeating electron gas in the crust, as well as the core of the
star which is assumed to be tightly coupled to the solid crust.

\subsection{Relative Rotation of the Vortices}
In order to determine the azimuthal component of the relative
velocity, $v_{\rm rel}$, or equivalently the relative rotational
frequency $\Delta \Omega \equiv | \Omega_{\rm m} - \Omega_{\rm
c}|$ between the unpinned vortices (rotating at a rate
$\Omega_{\rm m}$) and the crust, one might distinguish between two
distinct possibilities for the {\em initial conditions} upon
unpinning. When a vortex (segment) becomes unpinned it might be
expected to
\begin{itemize}
\item[{\em i)}] either, {\em initially tend} to maintain its
overall co-rotation with the pinned vortex lattice and the crust ,
as before unpinning, hence $\Delta \Omega =0$ {\em initially upon
unpinning} (ie. $ \Omega_{\rm m} = \Omega_{\rm L} = \Omega_{\rm
c}$, where $\Omega_{\rm L}$, defined earlier, is the rotation rate
of the pinned vortex lattice, in contrast to $\Omega_{\rm m}$ for
the temporary unpinned moveable vortices). This could arise due to
the general requirement for a locally uniform vortex distribution
imposed by the minimization of the free energy
(\markcite{SF68}Stauffer \& Fetter 1968), assuming that the
relaxation to the state of minimum energy of the system for the
new pinning conditions is achieved quickly enough compared to the
other timescales involved.

Thence, if the crust is not itself being acted upon by any
external torque, the situation may persist as the steady state,
while the superfluid keeps rotating at a different rate than its
container (and the vortices), keeping $\omega$ constant with time.
The unpinned vortices would be however under the influence of a
radial Magnus effect $(F_{\rm M})_r$, corresponding to the assumed
value of the lag $\omega$ (Eqs~3 \& 4). The tension of a vortex
line might be invoked as a possible source for counter balancing
the radial Magnus term, in the vortex equation of motion, for such
unpinned vortices having no radial motion.

If, on the other hand, the crust is itself being spun down by an
external torque, which is the case for a neutron star, a relative
azimuthal velocity could then develop, with the steady-state
magnitude (see \S~2.5, below)
\begin{eqnarray}
\Delta \Omega \sim  {\ N \over \ I_{\rm c} } \ \tau_v,
\end{eqnarray}
where $I_{\rm c}$ is the moment of inertia
of the crust (the rest of the star apart from the superfluid part
considered), and $N$ is the magnitude of the external torque
acting primarily on the crust. \item[{\em ii)}] Else, an unpinned
vortex might jump to a rotation frequency same as the superfluid,
instantaneously upon unpinning. Hence,
\begin{eqnarray}
\Delta \Omega  \sim  \Omega_{\rm s} - \Omega_{\rm L} \equiv
                  \omega.
\end{eqnarray}
The supporting argument for such an assumption would be the fact
that, in general, vortices are expected to move with the local
superfluid velocity; also a general requirement of the vortex
dynamics in the absence of external forces on the vortices
(\S~2.1). An instantaneous change of the rotational velocity is
indeed permitted for the vortices, being massless fluid
configurations, in the usual approximation of zero inertial mass
for a vortex (\markcite{son87}Sonin 1987; \markcite{bach83}Baym \&
Chandler 1983).

Thence, if the crust is not itself being acted upon by any
external torque ($N=0$), the superfluid would be spun down at the
expense of spinning {\em up} of the crust, and $\omega$ decreases
gradually. In the presence of negative external torque $N$,
however, $\omega$ may as well increase with time.
\end{itemize}
Either of the above two possibilities might provide a better
approximation depending on whether a vortex unpins as a whole
along its length, or only small segments of it are unpinned
randomly. For the superfluid in the crust of a neutron star,
simultaneous unpinning of a vortex as a whole must be ruled out,
given the huge number of the pinning centers (ie. the nuclei of
the solid crust) along each vortex (having a length of a km or
so); hence case {\em (i)} should be more probable. In contrast,
case {\em (ii)} might be the proper choice for the laboratory
experiments in which a vortex pins only at its end points
(\markcite{hed80}Hedge \& Glaberson 1980; \markcite{sch81}Schwarz
1981). Further theoretical work may indicate the extent to which
the (statistically averaged) motion of individual vortices could
deviate from a uniform local density, and distinguish between the
above alternative possibilities for the initial conditions of the
rotation rate of the vortices upon unpinning. The relaxation
timescale of the vortex array to the new conditions, in each case,
would be likewise relevant for making a decision. Also, the
distinct behavior of the superfluid spin-down, for $N=0$, in the
two cases might be possible to be tested, experimentally.

\subsection{General Two-Component Rotation}
An assumed  general model of a normal (non-superfluid) component
plus the ``crust'', with moments of inertia $I_{\rm n}$ and
$I_{\rm c}$, and rotation frequencies $\Omega_{\rm n}$ and
$\Omega_{\rm c}$, respectively, under the influence of an external
negative torque $- N$ acting primarily on the crust-component,
would obey the following dynamical relations (\markcite{Bayet69}
Baym et.~al. 1969)
\begin{eqnarray}
I_{\rm n} \dot\Omega_{\rm n} & = & I_{\rm c} \frac{\ \Omega_{\rm
c}
- \Omega_{\rm n}}{\tau_v}, \\
I_{\rm c} \dot\Omega_{\rm c} & = & -N -I_{\rm c} \frac{\
\Omega_{\rm c} - \Omega_{\rm n}}{ \ \tau_v},
\end{eqnarray}
where $I  =  I_{\rm c} + I_{\rm n}$, and $\tau_v$ is the
velocity-relaxation time for the dissipation of microscopic
relative motion between the constituent particles of the two
components. A solution of the two coupled equations indicate
exponential relaxations of the rotation frequencies $\Omega_{\rm
c}(t)$ and $ \Omega_{\rm n}(t)$, with time $t$. The exponential
time constant $\tau_{\rm D}$, referred to as the dynamical
coupling timescale of the system, is given as
\begin{eqnarray}
\tau_{\rm D} = {{\ I_{\rm n}}\over I} \ \tau_v.
\end{eqnarray}
In the case of a superfluid component, the relation between
$\tau_{\rm D}$ and $\tau_v$ would be in general different than
that in Eq.~12, as discussed below. Further, the steady-state
behavior inferred from the asymptotic solutions of Eqs~10 \& 11
indicate a relative rotation difference $\Delta \Omega_{\rm ss}$
such that
\begin{eqnarray}
\Delta \Omega_{\rm ss} = \Omega_{\rm n} - \Omega_{\rm c} =  -{{\
I} \over {\ I_{\rm c}}} \ \tau_{\rm D} \ \dot\Omega_\infty,
\end{eqnarray}
where \( \dot\Omega_\infty = - {{\ N} \over I} \) is the
steady-state spin-down rate of either component. The latter
relation (Eq.~13) is however expressing a general dynamical
relation, applicable also to the case of a superfluid component,
with the reservation that the relative rotation of the vortices
(not the superfluid) and the crust would be the relevant quantity.

\subsection{The Superfluid Dynamical Relaxation time}
In contrast to the above formulation of a two-component system,
the dynamical coupling time scale of a superfluid is associated
with the relaxation of its vortices to their new positions, in
response to the existing torque on the superfluid. The added
complexity is due to the fact that, unlike the particles of a
normal component, the relaxation of vortices involves both their
azimuthal as well radial displacements. Moreover, in the case of
random unpinning a further complication is that only a fraction of
the total vortices are effectively moving, at any given time. For
a pinned superfluid with a total number density of the vortices
$n_{\rm v}$, per unit area, random unpinning events at a rate
$\xi$ may result in a statistical population of free potentially
movable vortices, with a number density $n_{\rm m} = \xi n_{\rm
v}$ (Eq.~5), at any given time while $|\omega| < \omega_{\rm
crit}$. Likewise, looking at any given vortex over a large enough
time period (larger than the associated pinning/unpinning
intervals), it would move and take part in the relaxation process
for only a fraction $\xi$ of the time, and spends the rest of it,
$(1-\xi)$ fraction, as stationary pinned and decoupled. The drag
force on any unpinned moving vortex is nevertheless the same as in
the normal case when all of the vortices are free and mobile,
under the same assumed conditions for the scattering processes and
relative velocities (same $\tau_v$ and $v_{\rm rel}$). Also, the
instantaneous kinematic contribution of the vortices in the
superfluid spin frequency is the same irrespective of their
pinning/unpinning states. Therefore, the equation of motion of
each vortex, governing the time behavior of its radial
displacement between successive pinning events, would be exactly
the same as in the absence of any pinning (Eq.~14, below). A
superfluid rotational relaxation would nevertheless be achieved
via rearrangement of the (radial) positions of {\it all} vortices.
The distinction between a pinned subgroup and another unpinned is
meaningful only for the instantaneous considerations, and not for
a long term relaxation process. This would be further justified if
the vortices (being indistinguishable fluid entities) are required
to maintain a locally uniform density and more so if the time
between successive pinning/unpinnings for each vortex (being of
the order of the travel time between adjacent pinning sites, which
are the atomic nuclei in the solid crust of a neutron star) is
much shorter than the associated relaxation time. Thus the
vortices, under the assumed pinning conditions, take part in the
relaxation process {\it as a whole}, even though each undergoes an
intermittent cycle of movements and halts.

The relaxation time, in the absence of any pinning, is deduced
from a solution of the vortex equation of motion (Eq.~3) for the
radial $r_{\rm v}(t)$ and azimuthal $\phi_{\rm v}(t)$ components
of the vortex position in polar coordinates, as a function of time
$t$ (\markcite{as88} Alpar \& Sauls 1988;
\markcite{mj98}Jahan-Miri 1998):
\begin{eqnarray}
r_{\rm v}(t) & = & r_0 \left[ {{\Omega_{\rm s}}_0 \over \
{\Omega_{\rm c}}_0 } + \left( 1- {{\Omega_{\rm s}}_0 \over \
{\Omega_{\rm c}}_0 } \right) e^{-t/{\tau_{\rm D}}} \right]^{1/2}
\\ \phi_{\rm v}(t) & = & \phi_0 + {\Omega_{\rm c}}_0 t + K
\ln{\left( {r_{\rm v}(t) \over \ r_0 } \right) }
\end{eqnarray}
where 0-subscripts indicate initial values at $t=0$ corresponding
to an assumed departure from an earlier state of co-rotation of
the superfluid (vortices) and the crust, and $K = { \rho_{\rm s}
\kappa n_{\rm v} \over \ \rho_{\rm c}} \ {\tau_v}$. The relaxation
time $\tau_{\rm D}$ needed for the simultaneous re-adjustment of
the vortices in both radial and azimuthal directions in response
to the exiting torque on the superfluid, ie. the dynamical
coupling time scale, is given as
\begin{eqnarray}
\tau_{\rm D} & = & {\  K + {1 \over \ K} \over 2 {\Omega_{\rm c}}
\left( 1 + {I_{\rm s} \over \ I_{\rm c}} {{\Omega_{\rm s}} \over \
{\Omega_{\rm c}}} \right)},
\end{eqnarray}
where ${ \rho_{\rm s} \over \ \rho_{\rm c}} ={I_{\rm s} \over \
I_{\rm c}}$, and $\kappa n_{\rm v} = 2 \Omega_{\rm s} \approx 2
\Omega_{\rm c}$ have been used, omitting the zero subscripts.

In the case of pining, the radial position of each vortex changes
according to the same equation 14, between its successive pinned
states, followed by a halt in its motion until unpinning again.
For definiteness, we assume the typical time period $t_{\rm c}$
that any given vortex undergoes a pinning/unpinning cycle is much
shorter than the sought relaxation time scale of the system. This
should be the relevant limit for the case considered, given the
microscopic distances between pinning centers which set the the
order of magnitude of the typical distance that is travelled by an
unpinned vortex before re-pinning. This length scale together with
the typical relative (radial as well as azimuthal) velocities of
the vortices with respect to the crust will set the period $t_{\rm
c}$, for a given unpinning probability $\xi$. Thus, one needs to
do some averaging over successive movements and stationary states
of each vortex in order to infer an exponential-like time behavior
for its radial displacement, hence deducing a dynamical relaxation
time, comparable to the case of no pining. We try three different
averaging methods, which nevertheless give consistent results at
least for the relevant limiting cases.
\subsubsection{dynamical averaging}
As indicated, any given vortex is influenced, for a fraction $\xi$
of the time, by the {\em same} external force $\vec{F}_{\rm ext}$
as if there where no pinning, and zero azimuthal force in the rest
of its time. The time averaged motion of the vortex may be thus
determined, in the linear approximation, by a time averaged value
of the force, $\overline{\vec{F}}$. Assuming further that
$\vec{F}_{\rm ext}$ remains constant during the motion of the
vortex between its successive pinning states, one derives simply
\begin{eqnarray}
\overline{\vec{F}} & = &   \xi  \ \vec{F}_{\rm ext} \\
                  & = & \xi \  {\rho_{\rm c} \over n_{\rm v}} \
                        {\vec{v}_{\rm c} - \vec{v}_{\rm L}  \over
                        \tau_v},
\end{eqnarray}
for the effective value of the external force on each vortex, per
unit length, in the presence of pinning. The latter assumption of
the constant force is justified since it is being applied to a
time period much shorter than the relaxation time scale of the
system.

Alternatively, the equation of motion of the vortices (Eq.~3,
which applies to a single vortex, per unit length, as such) might
as well be integrated and averaged over radial distances (radial
shells) much larger than the microscopic distances between the
pinning sites, in the crust of a neutron star. This is justified
by the general requirement for a uniform local density of the
vortices, which supports the validity of a fluid dynamical
approach to the superfluid dynamics, in general
(\markcite{son87}Sonin 1987; \markcite{bach83}Baym \& Chandler
1983). Given the km size of the superfluid in the present case,
the integration volume would be populated by a large number of the
unpinned and pinned vortices, at any given time. Hence, for a
solution of the equation of motion, of the whole vortices within
an integration volume, one might as well think in terms of an
statistically averaged drag force. The averaging would obviously
give the same result as in Eqs~17-18, for the ``effective'' value
of the external force on each vortex, per unit length, in the
presence of pinning. The latter derivation of  Eqs~17-18 indeed
applies instantaneously, and dismisses with the earlier
restriction about the short term constancy of the drag force.

Solving the vortex equation of motion (Eq.~3), with
$\overline{\vec{F}}$ replaced for $\vec{F}_{\rm ext}$ therein, the
results would be similar to those in Eqs~14--16, except for the
timescale $\tau_{\rm D}$ which may be replaced by a corresponding
quantity $\tau_{\rm P}$ as the the dynamical coupling time scale
of the pinned superfluid:
\begin{eqnarray}
\tau_{\rm P} & = & {\  {K \over \xi} + {\xi \over \ K} \over
                   2 {\Omega_{\rm c}} \left( 1 + {I_{\rm s} \over
                   \ I_{\rm c}} {{\Omega_{\rm s}} \over \ {\Omega_{\rm c}}}
                   \right)} \\
             & \sim & {{I_{\rm c} \over \  I}} \left[{I_{\rm s} \over
                     \ I_{\rm c}} \ {\ \tau_v \over \xi} + {I_{\rm c} \over
                     \ I_{\rm s}} \ {\xi \over \ 4 \Omega_{\rm s}^2 \ \tau_v }
                     \right].
\end{eqnarray}
In the limit of $\tau_v >> {2 \pi \over \ \Omega_{\rm s}}$, which
is probably the appropriate limit for an application to the crust
of young neutron stars, this reduces to the approximate form
\begin{eqnarray}
\tau_{\rm P} \sim {\tau_{\rm D} \over \xi} \sim {{\ I_{\rm s}}
\over {\ I}} \ {\ \tau_v \over \xi},
\end{eqnarray}
also in agreement with the general results of the above
two-component model (Eq.~12), for $\xi=1$, as expected in that
limit where the effect of the vortex radial displacement may be
neglected thence vortex relaxation behaves as normal fluid,
approximately.
\subsubsection{kinematical averaging}
The dynamical relaxation time $\tau_{\rm P}$, in presence of
pinning, might be as well deduced from the time behavior of the
vortex motion, over many successive pinning/unpinning cycles. We
are again assuming $\tau_{\rm P} >> t_{\rm c}$, as argued above.
As depicted in Fig.~1a, the radial position of {\em any given}
vortex, in the pinned case, describes an exponential-like rise in
the radial-position--time diagram over a period $\xi t_{\rm c}$,
as predicted by Eq.~14 initially for the case of no pinning,
followed by a flat portion extended for another period of time $(1
-\xi) t_{\rm c}$. This pattern would be then repeated, with the
cycle time $t_{\rm c}$, until the final position is reached,
corresponding to an assumed final frequency of the superfluid. In
comparison to $\tau_{\rm D}$ which is defined as the time constant
associated with an exponential fit to the curve (ie., the function
in Eq.~14 ) described by each vortex in the case of no pinning,
$\tau_{\rm P}$ would be likewise the time constant associated with
an exponential fit to the whole curve representing the overall
motion of each vortex. As indicated earlier, for a long term
and/or steady state consideration as that of deducing a relaxation
time scale ($\tau_{\rm P} >> t_{\rm c}$) for a superfluid in
presence of pinning, {\em all} the vortices play the same and
equal role; the distinction between pinned and unpinned
populations is but an instantaneous fact. Hence, in the linear
approximation which makes it also possible to proceed further
analytically, simple geometrical considerations then give
(Fig.~1b)
\begin{eqnarray}
\tau_{\rm P} \sim {\tau_{\rm D} \over \xi},
\end{eqnarray}
also in agreement with the earlier approximate result as in
Eq.~21.
\subsection{The Rate}
The superfluid spin-down rate may be expressed, in its general
form, as (compare Eq.~13)
\begin{eqnarray}
\dot\Omega_{\rm s}  = {\ I_{\rm c} \over I } \ { \
                     \Delta\Omega_{\rm s} \over \tau_{\rm P}},
\end{eqnarray}
noticing that $\tau_{\rm P}$ is, by definition, the characteristic
time for the relaxation of a difference in the rotation frequency
$\Delta\Omega_{\rm s} \ (\equiv \Omega_{\rm s} - \Omega_{\rm c})$
between the superfluid and the container/crust. In the case of
pinning, however $\Delta\Omega_{\rm s}$ has to be replaced by the
relevant quantity $\Delta\Omega$, which was defined earlier (\S
2.4) as the difference in rotation frequency between the {\em
unpinned movable} vortices and the crust ($\Delta\Omega \ \equiv
\Omega_{\rm m} - \Omega_{\rm c})$. Obviously, the superfluid
relaxation would be sensitive to the relative rotation of the
crust with respect to the moveable vortices as the {\em only}
means for transmission of a torque. In other words,
$\dot\Omega_{\rm s} =0$ if and only if $\Delta\Omega = 0$. In
contrast, a steady state value of the lag between the superfluid
and the {\em pinned} vortices, implies $\Delta\Omega_{\rm s} \neq
0$, even though $\dot\Omega_{\rm s} = 0$, in the absence of the
external torque on the container/crust.

Substituting in Eq.~23, for $\tau_{\rm P}$ from Eq.~20, and
$\Delta\Omega$ (in place of $\Delta\Omega_{\rm s}$) from either
Eq.~8 or Eq.~9, the superfluid spin-down rate in presence of
random unpinning of the vortices is predicted to be
\begin{itemize}
\item {\bf case i)} if unpinned vortices {\em tend} to co-rotate
with the vortex lattice
\begin{eqnarray}
 \dot\Omega_{\rm s}  &=& { {\ N \tau_v} \over I_{\rm c}} \left[{I_{\rm s} \over \ I_{\rm
c}} \ {\ \tau_v \over \ \xi} + {I_{\rm c} \over \ I_{\rm s}} \
{\xi \over \  4 \Omega_{\rm s}^2 \ \tau_v } \right]^{-1}
\end{eqnarray}
\item {\bf case ii)} if unpinned vortices tend to co-rotate with
the bulk superfluid
\begin{eqnarray}
\dot\Omega_{\rm s}  &=& \omega \left[{I_{\rm s} \over \ I_{\rm
c}} \ {\ \tau_v \over \ \xi} + {I_{\rm c} \over \ I_{\rm s}} \
{\xi \over \  4 \Omega_{\rm s}^2 \ \tau_v } \right]^{-1}.
\end{eqnarray}
\end{itemize}
The corresponding average vortex radial velocity (Eq.~2) may be
written down as well, using the approximate limiting form of
$\tau_{\rm P}$ (Eq.~21, or Eq.~22),
\begin{eqnarray}
     v_r \sim  \begin{array}{ll}
                \frac{r}{2 \Omega_{\rm s}} { {\ N } \over I_{\rm s}} \ \xi &
                \ \ \ \ \ \mbox{case ($i$)} \\
                             \frac{r}{2 \Omega_{\rm s}}
         \frac{I_{\rm c}}{I_{\rm s}} \frac{\omega}{\tau_v} \ \xi &
               \ \ \ \ \  \mbox{case ($ii$)}.
                            \end{array}
\end{eqnarray}
The dependence on $N$, $\omega$, and $\tau_v$, even in these
simplified forms of the relation, represents the very dependence
of the superfluid spin down rate on the {\em instantaneous torque
exerted on the superfluid by its environment/container}. As a
specific manifestation of this dependence, the sign of $v_r$, that
is the sign of the change in the superfluid spin rate, is
determined by that of $N$, or $\omega$, in either cases. As
expected (\S~2.4) Eq.~26 also confirms that, in the absence of
external torque $N$ {\em on} the the superfluid {\em container},
the pinned superfluid may either retain its rate or else come to a
state of co-rotation with the container, depending on the two
possibilities considered for the rotation rate of the vortices
upon unpinning. The uncertainties in the (micro)physics of
individual vortex motion, within a vortex lattice, prevent from
deciding between the two cases. However, the predicted distinct
behaviors, for the case of $N=0$, might be used in possible
laboratory experiments as a clue to distinguish between the two
cases. As a further confirmation, Eq.~26 (case ii) reduces, as it
should, to the correct form expected in the absence of pinning
(\markcite{adam85} Adams, Cieplak \& Glaberson 1985;
\markcite{as88} Alpar \& Sauls 1988; \markcite{mj98}Jahan-Miri
1998), for the limiting case of $\xi =1$ corresponding to values
of $|\omega| \geq \omega_{\rm crit}$, when the Magnus effect
prevents (even temporary) pinning to be realized.

The above prediction (Eqs~24 or 25) for the superfluid spin-down
rate, driven by random unpinning events with a given probability
$\xi$, is fundamentally different than the earlier predictions
(\markcite{alet84} Alpar et.~al. 1984; \markcite{MJ05a}Jahan-Miri
2005a). The correct dependence on the dynamically relevant
quantities $N$, $\omega$ , $\tau_v$, and $\xi$ assures a true and
instantaneous dependence of the superfluid spin-down rate
$\dot\Omega_{\rm s}$ (or equivalently $v_r$) on the sing and
magnitude of the actual torque transmitted between the superfluid
and its container/environvemnt (the crust). It may be noted that
even though the steady-state magnitude of $\omega$ would be set by
other dynamically independent quantities, however for a transient
post-glitch relaxation which is our prime objective here it is
indeed an independent evolving quantity, initially determined by
the glitch. The opposite dependence on $\xi$ in the two terms at
the right hand side of Eq.~20 (appearing also in Eqs~24 or 25 ) is
interesting, and resembles the similar behavior of the relaxation
time $\tau_v$. The new prediction reduces to an earlier reported
estimate (\markcite{MJ05a}Jahan-Miri 2005a), only in the
approximate form, as in Eq.~26, for the limiting cases indicated
(with a correction for the case {\em i} therein).

For a quantitative evaluation of the efficiency of the spinning
down of a superfluid through random unpinning of its pinned
vortices, an order of magnitude estimate of the maximum spin-down
rate predicted by the present model (Eqs~24 or 25) may be given,
as applicable to the crust of neutron stars. The spin-down rate
indeed depends on the instantaneous number of the unpinned
vortices, as determined by the unpinning probability function
$\xi(\omega)$. The maximum spin-down rate would be achieved for
values of $\xi \sim 1$, corresponding to $\omega \sim \omega_{\rm
crit}$. Adopting a set of parameter values applicable to
post-glitch relaxations in young neutron stars, such as $r \sim
10^6 \ {\rm cm}$, $\Omega_{\rm s} \sim 10^2 \ {\rm rad~s}^{-1}$,
$N/I \sim 10^{-10} \ {\rm rad~s}^{-2}$, $I_{\rm s}/I \sim 0.02$,
$\omega_{\rm crit} \sim 10^{-2} \ {\rm rad~s}^{-1}$, $\tau_v \geq
10 \ {\rm s}$, the (averaged) radial velocity of the vortices
could be (Eq.~26)
\begin{eqnarray}
     v_r  \sim  \begin{array}{ll}
    10^{-4} \ {\rm cm \ s}^{-1}   &
         \ \ \ \ \ \mbox{case ($i$)} \\
    10^{3} \ {\rm cm \ s}^{-1}&
               \ \ \ \ \  \mbox{case ($ii$)},
                            \end{array}
\end{eqnarray}
corresponding to the superfluid spin-down rates (Eq.~2)
\begin{eqnarray}
  \dot\Omega_{\rm s}   \sim  \begin{array}{ll}
    10^{-8} \ {\rm rad \ s}^{-2}   &
         \ \ \ \ \ \mbox{case ($i$)} \\
    10^{-1} \ {\rm rad \ s}^{-2}&
               \ \ \ \ \  \mbox{case ($ii$)}.
                            \end{array}
\end{eqnarray}
The parameter values used above are indeed case dependent to a
large extent, and also the exact expression for $\tau_{\rm P}$
(Eq.~20) should be used for a more accurate quantitative estimate.
The much larger uncertainty lies however in deciding between the
two cases indicated. Nevertheless, the predicted maximum rate,
even for that of case ($i$), is seen to be generally (much) larger
than the observed spin-down rates of the radio pulsars, by at
least one order of magnitude for the Crab and much so in the case
of other pulsars. Therefore, the pinned superfluid in the crust
may as well spin down at the same steady-state rate as the rest of
the star, through random unpinning events with an associated value
of $\xi \lesssim 0.1$ or much smaller, maintaining a rotational
lag smaller than the critical lag value.

The referee is acknowledged gratefully for his/her extended and
valuable comments/corrections leading to much improvement of the
manuscript. This work was supported by a grant from the Research
Committee of Shiraz University.

\clearpage \figcaption{{\bf a)} A sketch of the radial
displacement, $r_v$, of any given vortex (in a given shell) versus
time, $t$, during a rotational relaxation of the superfluid. A
superfluid relaxation to an assumed final rotation frequency
corresponds to a change in the vortex density, hence to certain
radial displacement $r_{\rm v}(\infty)$ of each vortex. In the
absence of any pinning ({\em dotted line}) $r_{\rm v}(t)$ is an
exponential-like function, as in Eq.~14, with an associated time
constant $\tau_{\rm D}$, defined as the dynamical time scale of
the superfluid. In presence of pinning with an assumed rate $\xi$
of random unpinning, {\em all} vortices pin/unpin intermittently
with a cycle time $t_{\rm c}$. Each vortex describes the same
function $r_{\rm v}(t)$ as in Eq.~14, however for only a time
period $\xi \ t_{\rm c}$, then it stops and stays pinned for
another time period $(1 - \xi) \ t_{\rm c}$. The same cycle of
motion/halt is repeated until the final displacement $r_{\rm
v}(\infty)$ is reached, as required by the assumed final frequency
of the superfluid in a given relaxation. {\bf b)} Same as (a), but
in the linear approximation, which also makes it possible to
deduce, analytically, a corresponding dynamical time scale
$\tau_{\rm P}$ for the  pinned superfluid. The {\em thick solid
line} is a convenient linear fit to the linear approximation of
the actual time behavior ({\em thin solid line}) of the
displacement of each vortex in the presence of pinning and random
unpinning, while the {\em dotted line} is the linear approximation
for the case of no pinning. Note that $\xi$ has been largely
exaggerated, as compared to its typical expected values, for the
demonstration.} \label{f1} \clearpage
\begin{figure}
\plotone{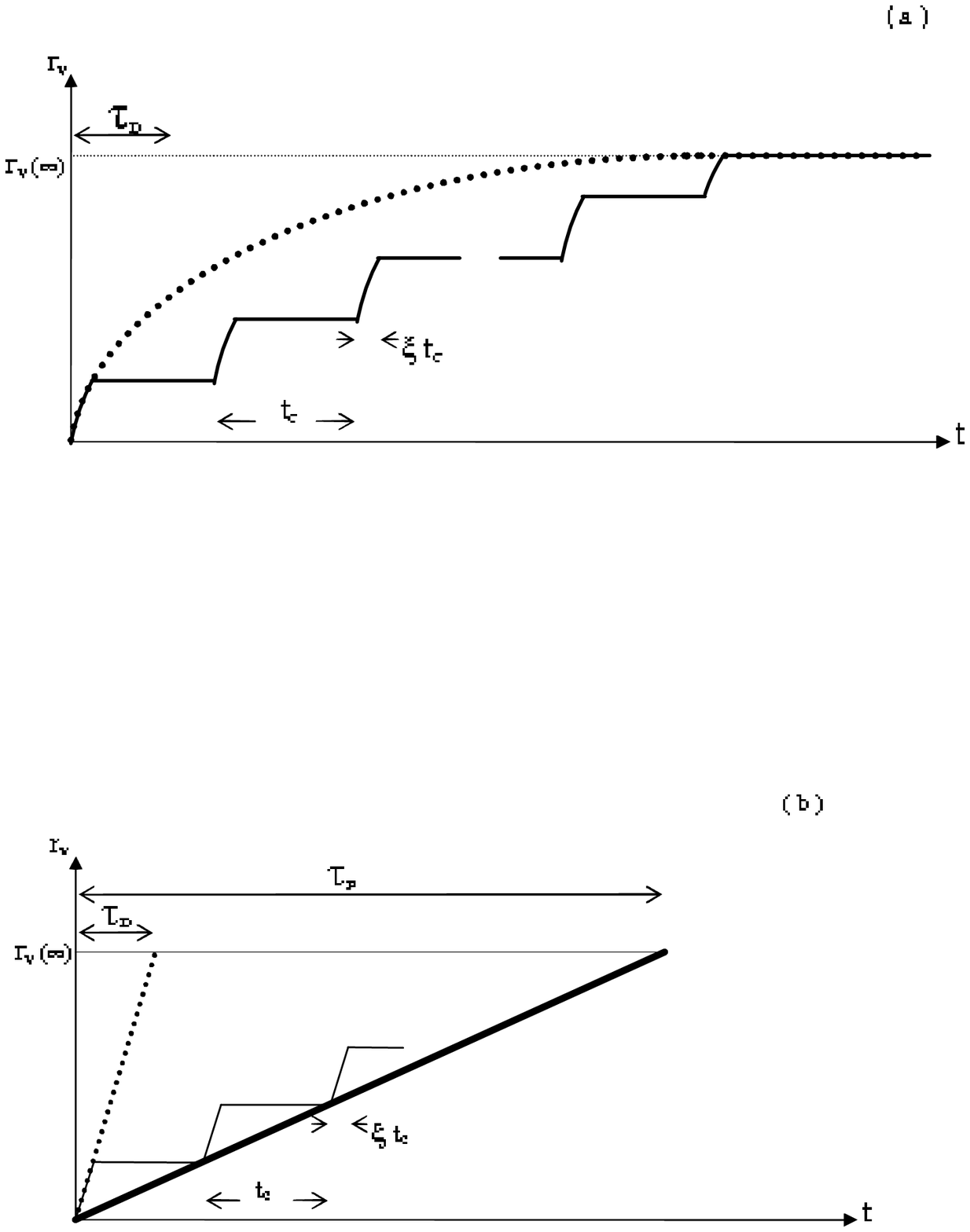}
\end{figure}

\end{document}